\documentclass{article} 

\pdfoutput=1

\usepackage{amsmath,amssymb,amsthm,graphicx}
\graphicspath{{code/}}
\usepackage{algorithm}
\usepackage{algorithmic}
\usepackage[top=2cm, bottom=2cm, left=2.5cm, right=2.5cm]{geometry}
\usepackage[T1]{fontenc}

\DeclareMathOperator*{\argmax}{arg\,max}
\DeclareMathOperator*{\argmin}{arg\,min}
\DeclareMathOperator*{\var}{var}
\DeclareMathOperator*{\cov}{cov}

\newcommand{\E}{\mathbf{E}}
\newcommand{\data}{\mathcal{D}}

\newcommand{\fmin}{f^\ast}
\newcommand{\xmin}{x^\ast}

\newcommand{\fminest}{\hat{f}^\ast}
\newcommand{\xminest}{\hat{x}^\ast}
\newcommand{\xdomain}{\mathcal{X}}

\begin{document}
\title{Active Bayesian Optimization: Minimizing Minimizer Entropy}
\author{
Il Memming Park\\
Center for Perceptual Systems\\
University of Texas at Austin\\
Austin, TX 78712, USA\\
\texttt{memming@austin.utexas.edu}\\
\and
Marcel Nassar, Mijung Park\\
Electrical and Computer Engineering\\
University of Texas at Austin\\
Austin, TX 78712, USA\\
\texttt{\{nassar.marcel, mjpark\}@mail.utexas.edu}
}
\maketitle

\begin{abstract}
The ultimate goal of optimization is to find the minimizer of a target function.
However, typical criteria for active optimization often ignore the uncertainty about the minimizer.
We propose a novel criterion for global optimization and an associated sequential active learning strategy using Gaussian processes.
Our criterion is the reduction of uncertainty in the posterior distribution of the function minimizer.
It can also flexibly incorporate multiple global minimizers.
We implement a tractable approximation of the criterion and demonstrate that it obtains the global minimizer accurately compared to conventional Bayesian optimization criteria.
\end{abstract}

\section{Introduction}
Exploring an unknown parameter space in search for globally optimal solution can be quite costly.
The aim of active optimization is to carefully choose where to sample in order to reduce the number of sample acquisitions; hence, reduce their cost~\cite{Jones1998}.

While possible, learning the response surface $f$ followed by a search for the minimizer is typically wasteful as not all regions of the response surface are of  interest; we do not need the details of the response surface in regions far from the optimum.
Under the active Bayesian optimization framework, the goal is to query the oracle, potentially noisy, as few times as possible while quickly gaining knowledge of the minimizer $x^\ast = \argmin_x f(x)$.
Prior works mostly focus on finding $\xmin$ by obtaining the function minimum $\fmin$ \cite{Kushner1964,Mockus1982,Elder1993dissertation,LizotteDissertation,Osborne2009,Gramacy2010,Srinivas2010}.
Such criteria will drive the sampling procedure towards improving the estimate of $\fmin$ and providing an estimate of $\xmin$ as a consequence.
Since $\fmin$ is unique while $\xmin$ might not be, such approaches often discard potential minimizers.
In design problems with cost constraints, this could lead to discarding viable and cost-effective solutions. %

In this paper, we use a acquisition criterion that maximizes the information gain about the minimizer, 
or equivalently \textit{minimizes minimizer entropy} (MME)~\cite{Villemonteix2008,Henning2011}.
MME provides a balance between exploration and exploitation that is tailored specifically for finding the minimizers of global optimization problems. 
As a result, the MME samples densely around potential minimizers, and sparsely in the other region of the input space~(Fig.~\ref{fig:toy} and \ref{fig:2D}). 
Furthermore, since a global map of potential minimizers is maintained, MME enables us to obtain multiple global minimizers.

\section{Optimization Framework} \label{sec:OpFrm}

We consider optimization target that is a continuous real-valued function $f:\mathcal{X} \mapsto \mathbb{R}$, where $\mathcal{X} \subset \mathbb{R}^d$ is bounded.
Furthermore, we assume $f$ has a unique minimizer $\xmin = \argmin_{x \in \xdomain} f\left(x\right) $ (an assumption relaxed later)
and that each observation is noisy; \textit{i.e.} $y|f,x \sim \mathcal{N}(f(x), \sigma^2)$.
The objective of the optimization is to find the function's minimizer $\xmin$ and its corresponding minimum $\fmin = \min_{x \in \xdomain} f(x)$.%

The Bayesian optimization framework has been proposed to arrive to an $\epsilon$-close solution in a \textit{sub-exponential} number of function evaluations on average~\cite{Kushner1964,Mockus1982,LizotteDissertation}.
While it is possible to devise strategies seeking the jointly-optimal $n$ samples, the computational cost is often prohibitive~\cite{Kushner1964,Mockus1982,Osborne2009}; hence, a greedy (or a one-step lookahead) sequential approach is typically used where the next sample is chosen according to an acquisition criterion.
Popular acquisition criteria are summarized in the table below:
\begin{center}
\begin{tabular}{c|c|p{7.2cm}}
Criterion & $x_{n+1}=\argmax_{x\in \xdomain}$ of & Description \\
\hline\hline
Kushner \cite{Kushner1964} & $ \Pr \left( f\left(x\right) < \fminest_n - \epsilon \right)$ & 
Samples the point with the highest probability of lying below the current minimum estimate. \\
\hline
Mockus~\cite{Mockus1982} & $\E \left\{ \left(f\left(x\right) - \fminest_n - \epsilon \right)_+\right\} $ & Samples the point with the largest expected improvement over the current minimum estimate.\\
\hline
\end{tabular}
\end{center}

\section{Proposed Acquisition Criterion: The MME Criterion}
\label{sec:ProposedAcquisitionCriterion}

The Bayesian  framework, when applied to functional estimation, defines a prior $p\left( f \right)$ over the functions $f$ and a corresponding posterior $p(f|\data_n)$ after $n$ observations.
Statistical inference on $\xmin$ requires the posterior $p\left(\xmin|\data_n \right)$.
The minimizer  $\xmin$  relates deterministically to $f$  through the highly nonlinear ``$\argmin$'' operation; hence, it is intractable to compute $p\left( \xmin | \data_n\right)$ from the  posterior $p\left(f|\data_n \right)$.
In particular, 
consider the set of points for which the function values are close to the optimum, $A = \{x \in \xdomain: f^\ast + \epsilon \geq f(x) | \data_n \}$ for a small $\epsilon > 0$.
$A$ may not be localized even for a smooth true $f$ since the function could have multiple disjoint $\epsilon$-close optimum regions (possibly due to multiple optimizers), making the minimizer distribution $\xmin|\data_n$ often quite complex.
Therefore, in this paper, we propose utilizing the inference on $f|\data_n$ as an intermediate step to learn $\xmin|\data_n$ more  efficiently by focusing the sampling on the regions of $\xdomain$ that contribute the most information about the minimizer.

Let $\xmin_n = x^*|\mathcal{D}_n$ be the random variable $x^*$ representing the minimizer conditioned on $n$ observations.
Our proposed criterion MME minimizes the minimizer entropy  $H(\xmin)$, where $H(\cdot)$ denotes the entropy functional.
In this paper, we focus on a sequential sampling scheme where we seek the next point $x_{n+1}$ that minimizes the entropy of the minimizer given the additional sample $\left(x_{n+1}, y_{n+1} \right)$. %
Thus, the next sample point is given by: %
\begin{equation}\label{eq:memmingObjective}
    \argmin_{x_{n+1}} H\left(x^*_{n+1} \right) = \argmin_{x_{n+1}} \E_{\mathbf{y_{n+1}}} \left[ H\left(\xmin|\data_n, \left(x_{n+1},\mathbf{y_{n+1}}\right) \right) \right].
\end{equation}
A straightforward evaluation of \eqref{eq:memmingObjective} requires the computation of %
\begin{equation}\label{eq:minimizerPosterior}
p\left(x^* |\data_{n+1} \right) = 
\int p(x^*|f) p\left(f|\data_{n+1}\right) \mathrm{d}f
=
\int \delta\left(x^* - \argmin_x f(x)\right) p\left(f|\data_{n+1}\right)
\mathrm{d}f.
\end{equation}

Since direct evaluation of \eqref{eq:minimizerPosterior} is intractable in general, we develop a more tractable approximation.
In this paper, we utilize the widely used Gaussian process framework for $f|\data_{n+1}$ ~\cite{Jones1998,LizotteDissertation,Osborne2009,Rasmussen2005}.
The minimizer's posterior in \eqref{eq:minimizerPosterior} can be pointwise bounded as follows
\begin{align}\label{eq:minimizerPosteriorBound}
p\left(\xmin = x | \data_{n+1} \right) =
    p\left(f(x) \leq f\left(x' \right),
    \forall x' \in \xdomain | \data_{n+1}\right)
\leq 
    p\left(f(x) \leq \fminest  | \data_{n+1}\right) 
\end{align}
where $\fminest = f(\xminest)$ is our current estimate of the minimum. The equality results from the definition of the minimizer, while the inequality is due to the fact that $f(x) \leq f\left(x' \right),\forall x' \in \xdomain$ implies $f(x) \leq f(\xminest)$.
The upper bound in \eqref{eq:minimizerPosteriorBound} is equal to $g_{n+1}\left(x \right)$ where 
\begin{align}
g_{n}(x) = p\left( f(x) \leq f(\xminest) | \data_{n} \right)
= \Phi\left( 
    \frac{\E[f(\xminest)] - \E[f(x)]}
    {\sqrt{\var{\left[f(\xminest)\right]} 
    + \var{\left[f(x)\right]}-2\cov{\left[f(x),f(\xminest)\right]}}}
\right)
\propto
\tilde{f_{n}}(x)
\label{eq:proxy}
\end{align}
where $\Phi$ denotes normal cdf, and the posterior means, variances, and covariances can be found in \cite{Rasmussen2005}.
Finally, we normalize $g_n(x)$ and use it as a proxy $\tilde{f_n}(x) \approx p(x^\ast = x|\data_n)$.

\begin{figure}[tb]
    \centering
    \includegraphics[width=\textwidth]{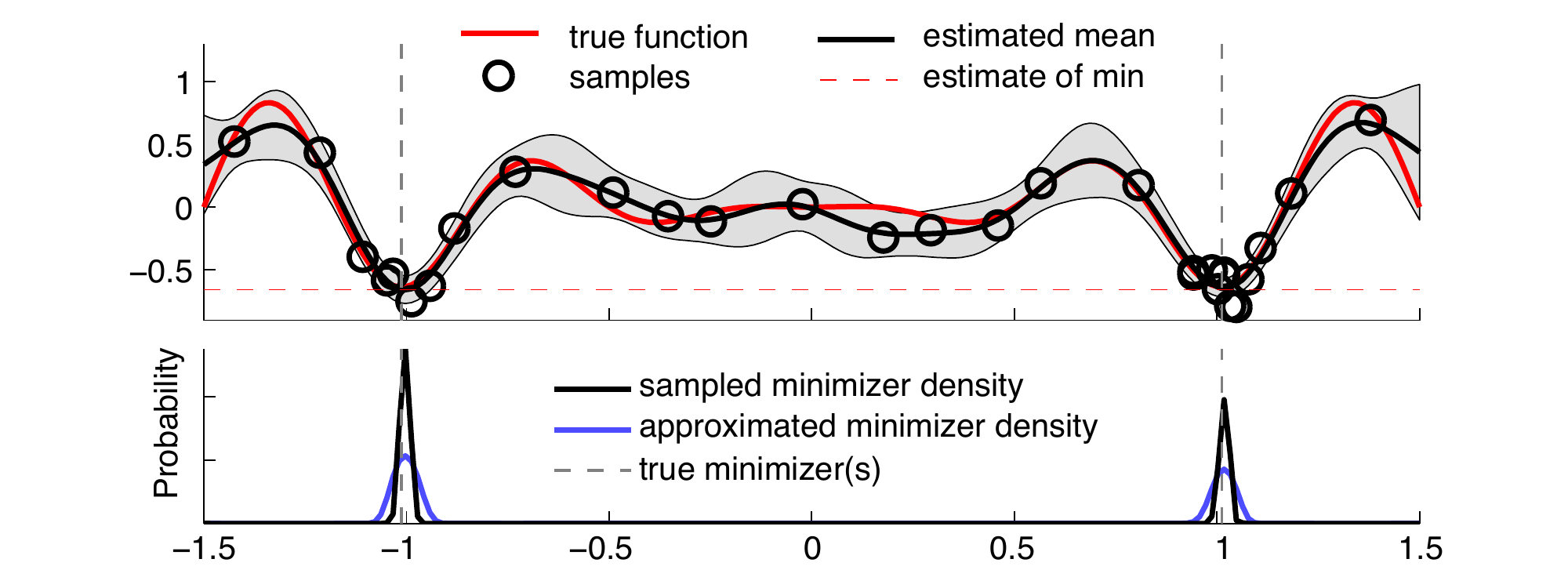}
    \caption{
    Illustration of minimizer distribution of $f(x) = (-e^{-x^2} + 1) \cos(3 \pi x)$ in $[-1.5, 1.5]$ with additive Gaussian noise (variance $(0.1)^2$).
    (Top) True target function $f$ and estimated posterior distribution via GP from 20 samples using MME.
    (Bottom) Minimizer distribution estimated from random samples of GP, and approximation by~\eqref{eq:proxy}.
    }
    \label{fig:toy}
\end{figure}

Notice that the approximation step leads to a broader distribution with respect to the true posterior $\xmin|\data_n$; hence, the  resulting entropy is an upper bound as well.
Moreover, in the noiseless case, the two distributions converge (for functions with unique minimizer); \textit{i.e.}, when the posterior $f|\data_n$ approaches to the true $f$, which is a delta function at the minimizer.
A natural advantage of this approximation is that it generalizes to functions with multiple global minimizers through the multi-modal $\tilde{f}(x)$ (see \figurename~\ref{fig:toy}). However, having multiple minimizers leads to  ambiguity in the posterior covariance term, $\cov{\left[f(x),f(\xminest)\right]}$, in \eqref{eq:proxy}. In this case, we treat $f(x)$ and $f(\xminest)$ as independent and remove the covariance term from \eqref{eq:proxy}.

\textbf{Implementation:}
For each sample acquisition, we have to estimate \eqref{eq:memmingObjective}.
This requires an expectation over $y_{n+1}$ for each candidate $x_{n+1}$; we use a Monte Carlo approach sampling $y_{n+1}|\data_n, x_{n+1}$ under the prior.
Given each $y_{n+1}$, we use the approximation~\eqref{eq:proxy} and evaluate the entropy~\eqref{eq:memmingObjective}.
This is done for each candidate on a grid, and the candidate that minimizes the criterion is chosen.
An alternative to sampling $y_{n+1}$, which can be costly, is to further approximate the expected posterior entropy by assuming the posterior mean function remains constant -- we refer to the algorithm with this extra assumption the ``fast'' version.

GP requires selection of kernels and associated hyperparameters.
Choosing a good hyperparameter is critical for good small sample performance, and convergence to global solution.
After acquiring each sample, we use evidence optimization to infer hyperparamters (including $\sigma^2$)~\cite{Rasmussen2005,Frazier2009}.
For the examples in the result section, we used isometric squared exponential kernel with two hyperparameters~\cite{Rasmussen2005}, and a constant mean function (1 hyperparameter).

\section{Results}
\subsection{1D toy example}
\begin{figure}[tb]
    \centering
    \includegraphics[width=\textwidth]{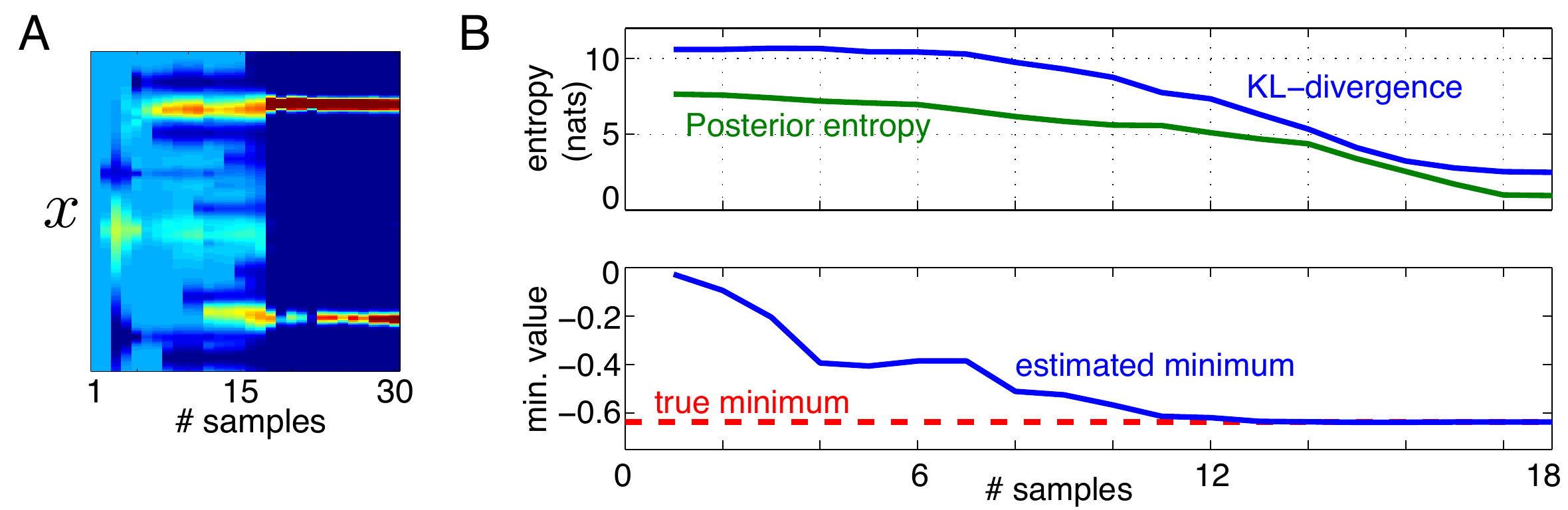}
    \caption{
	Convergence of the posterior minimizer distribution.
	The target function is identical to Fig.~\ref{fig:toy} without noise.
	\textbf{A:} Evolution of posterior distribution with the number of observations.
	Each vertical slice is colored density (transformed to enhance visualization).
	Note that it converges to two peaks corresponding to true minimizers.
	\textbf{B:} KL-divergence between the true minimizer measure and the posterior minimizer distribution, and the entropy of the minimizer distribution (top).
	Estimated function minima (bottom).
	Each trace is a median of 41 Monte Carlo runs.
    }
    \label{fig:1D:convergence}
\end{figure}

To illustrate the main ideas, we demonstrate the algorithm on a 1D function under additive Gaussian noise (\figurename~\ref{fig:toy}). The 1D function has two global minima and two local minima; therefore, the minimizer distribution is multi-modal. \figurename~\ref{fig:toy} shows the sampling distribution of the minimizer and our approximation of it. As expected, it has two peaks corresponding to the two global minima. Furthermore, \figurename~\ref{fig:1D:convergence} shows the evolution of the minimizer's posterior and its convergence to the sharp bimodal form given in \figurename~\ref{fig:toy}.

\subsection{2D examples}
We compare our criterion against the  popular criterion proposed by Mockus \cite{Mockus1982}, also called \textit{Maximum Expected Improvement} (MEI), on two 2D test functions with a noise variance of $(0.1)^2$:  Hosaki function (1 local, 1 global minimum)~\cite{Elder1993dissertation,Bekey1974}, and the Dixon-Szeg\H{o} 6-hump camel test function (2 local, 2 global minimum)~\cite{Dixon1978a}. In addition, to illustrate the effectiveness the Bayesian Optimization framework, we compare against the state-of-the-art active response surface method proposed by Krause \textit{et al.}~\cite{Krause2008}.
All algorithms were applied under the same prior and hyperparameter selection procedure with the only difference being the acquisition criterion.
Both MEI and the response surface approach require a relatively good (initial) estimate of the hyperparameters, therefore we initialize them with $10$ random samples.
Also, since the response surface method only works when the candidate set is finite, we restrict the sampling to be on a $15\!\times\!15$ grid for all algorithms.
\begin{figure}[t]
\centering
\includegraphics[width=1\linewidth]{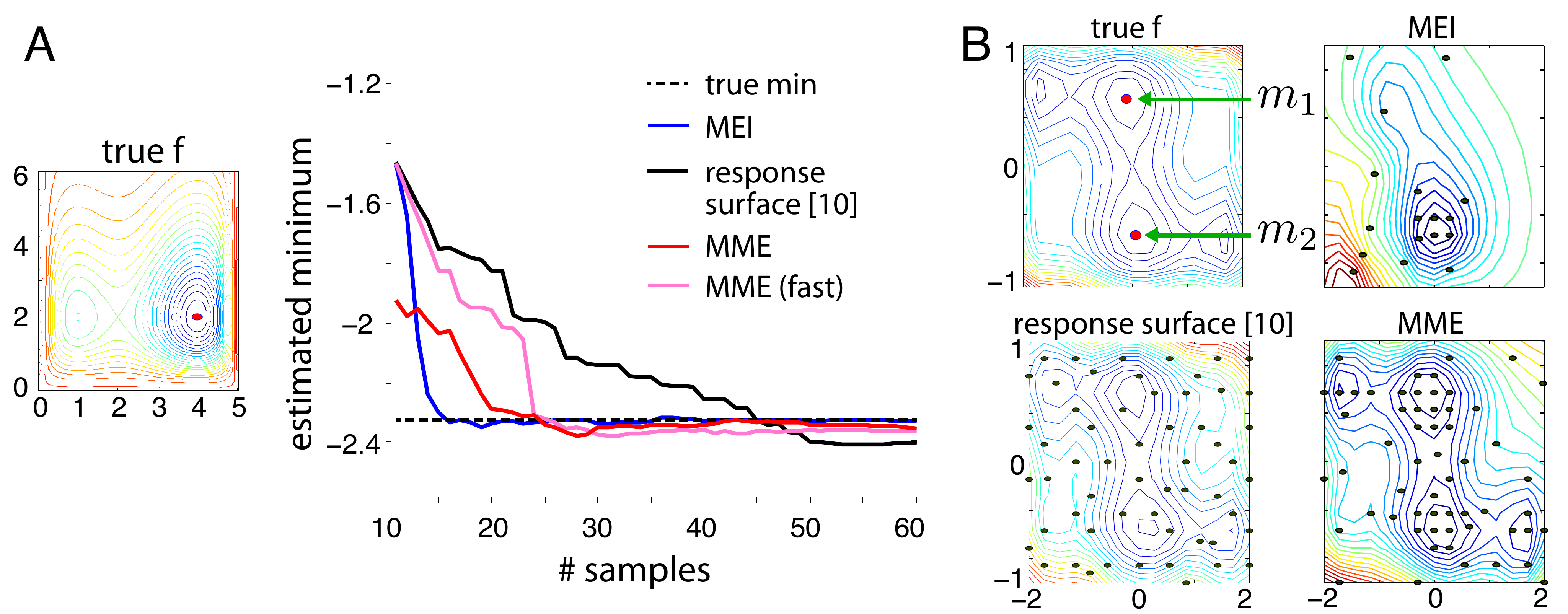}
 \caption{Comparison of acquisition criteria for 2D functions.
{\bf A}: Evolution of estimated minimum as a function of number of samples observed for Hosaki function (contour).
{\bf B}: 50 selected samples (black dots) for Dixon-Szeg\H{o} 6-hump camal function with two global minima (red dots).
}
\label{fig:2D}
\end{figure}
Fig.~\ref{fig:2D}A shows the convergence for the Hosaki function in terms of median of the estimated minimum function values obtained by each method from the $20$ repetitions.
Both MME and MEI performed well, while the response surface method has slower convergence and underestimates the correct minimum value due to the inaccurate estimation of $\sigma^2$.
For the Dixon-Szeg\H{o} test function, the MEI constantly drew samples (black dots in Fig.~\ref{fig:2D}B) near one of the global minimizers, and thus failed to find the other minimum.
The response surface method drew samples from all over the space and found the minimizers correctly, but the value of the minima were not as accurate.
On the other hand, MME found the correct minimizers and accurate minimum values compared to the other two methods.
The estimated minimum values at the global minimizers are shown in the table below.
\begin{center}
\centering
\begin{tabular}{|c|c|c|c|c|}
\hline
& true & M & K & MME \\
\hline
$m_1$ & -0.999 & 1.185 & -1.202 & -0.960 \\
\hline
$m_2$ & -0.999& -0.975 & -0.791 & -1.020 \\
\hline
\end{tabular}
\end{center}

\section{Discussion}
We proposed an information theoretic active optimization criterion by focusing on learning the minimizer distribution.
The problem of optimizing an unknown function is transformed to minimization of estimated entropy of minimizer obtained by Gaussian process.
We plan to improve computational complexity and approximation accuracy in future work.

\bibliographystyle{hunsrt}
\bibliography{references/ao}

\end{document}